\documentclass[twocolumn,shortnote]{jpsj2} 
\title{ Conventional $s$-Wave Superconductivity in Noncentrosymmetric Ir$_2$Ga$_9$: $^{71}$Ga-NQR Evidence}
\author{Atsushi \textsc{Harada}\thanks{Present address: National Institute for Materials Science, Ibaraki 305-0003, Japan.  E-mail address: HARADA.Atsushi@nims.go.jp}, Nobukatsu \textsc{Tamura}, Hidekazu \textsc{Mukuda}, Yoshio \textsc{Kitaoka}, Kouhei \textsc{Wakui}$^{1}$, Satoshi \textsc{Akutagawa}$^{1}$, and Jun \textsc{Akimitsu}$^{1}$}

\inst{Department of Materials Engineering Science, Osaka University, Toyonaka, Osaka 560-8531, Japan \\
$^{1}$Department of Physics and Mathematics, Aoyama-Gakuin University, Sagamihara, Kanagawa 229-8558, Japan\\ }

\recdate{\today}
\kword{noncentrosymmetry, superconductivity, Ir$_2$Ga$_9$, NQR }

\newpage
\begin{document}
\maketitle

Superconductivity without spatial inversion symmetry has begun to attract much attention since the discovery of the superconductor CePt$_3$Si \cite{Bauer}. It is presumed that the absence of spatial inversion symmetry causes an admixture between even and odd parities of superconducting pairing due to antisymmetric spin-orbit coupling (ASOC) \cite{Frigeri}. However, there has been no clear evidence of the admixture thus far, and hence the spatial noncentrosymmetric effect on superconductivity remains experimentally controversial. Recently, ASOC has also been discussed in transition-metal superconductors. In this system, we can consider ASOC without strongly correlated electron systems, which would be appropriate for tracing the effect of ASOC. Interestingly, in Li$_2$Pt$_3$B, it was reported that unconventional superconductivity with a line node and a spin-triplet pairing state emerge, which are ascribed to ASOC \cite{Yuan,Nishiyama}. These reports suggest that ASOC can induce a novel superconducting state even though there is no strong electron correlation.  

The intermetallic binary Ir$_2$Ga$_9$ does not possess spatial inversion symmetry and becomes superconducting below $T_{\rm c}=2.2$\,K \cite{Shibayama,Wakui}. Specific heat and resistivity measurements showed that the compound is a weak-coupling BCS superconductor with an isotropic gap, and located near the boundary between type-I and type-I\hspace{-.1em}I superconductivities with an upper critical field $H_{\rm c2}\sim 150$\,Oe \cite{Shibayama,Wakui}. In this letter, we report on the characteristics of superconductivity in noncentrosymmetric Ir$_2$Ga$_9$ probed by $^{71}$Ga-nuclear-quadrupole-resonance (NQR) measurement at a zero field ($H=0$). The spatial noncentrosymmetric effect on the superconducting state is discussed. 

A single crystal of Ir$_2$Ga$_9$ was grown by the Ga flux method \cite{Wakui}. Powder X-ray diffraction indicated that the compound forms in the primitive monoclinic Rh$_2$Ga$_9$ type structure \cite{Wakui,Bostrom}.
The sample of Ir$_2$Ga$_9$ was crushed into coarse powder for NQR measurement to allow the penetration of the rf field. The NQR measurement was performed by the conventional spin-echo method in the temperature ($T$) range of 1-280\,K. $T_1$ was measured at $f\sim 26.45$\,MHz, which was the 1$\nu_{\rm Q}$($\pm1/2 \leftrightarrow \pm3/2$) transition of $^{71}$Ga ($I=3/2$), as shown in Fig.~1. The NQR spectrum  ensures the quality of the sample, because the full width at half maximum (FWHM) is as small as 54 kHz, as shown in Fig.~1.  The NQR signal of the isotope $^{69}$Ga ($I=3/2$) was confirmed at $f\sim 42.08$\,MHz, which was consistent with the ratio $^{69}Q/^{71}Q \sim$\,1.59 ($Q$ is the quadrupole moment) because NQR frequency is proportional to $Q$. 
%
\begin{figure}[h]
\centering
\includegraphics[width=7.2cm]{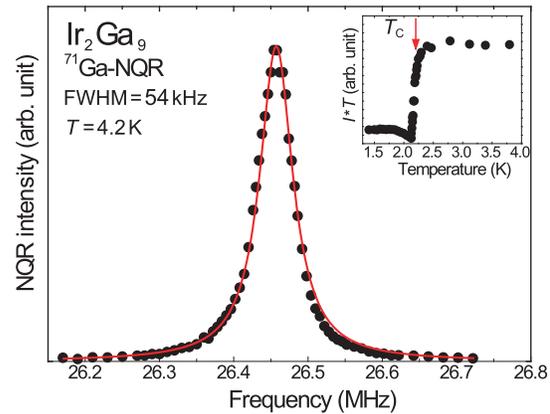}
\caption[]{(Color online) $^{71}$Ga-NQR spectrum of Ir$_2$Ga$_9$ at $T=4.2$\,K. The inset shows the $T$ dependence of $I\times T$. Here, $I$ is the NQR intensity.}
\label{NQRspectra}
\end{figure}

%
\begin{figure}[h]
\centering
\includegraphics[width=7.5cm]{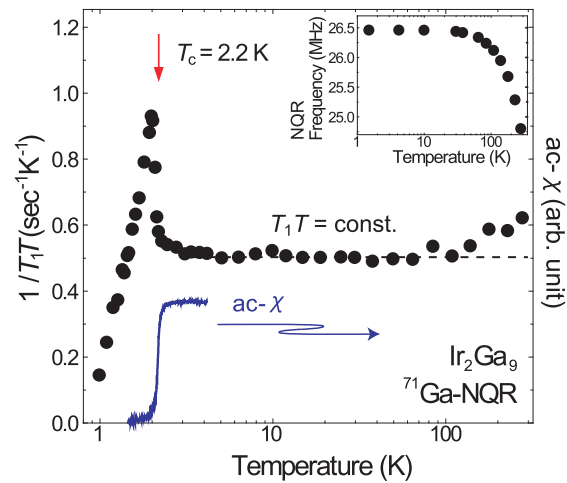}
\caption[]{(Color online) Temperature dependence of $1/T_1T$ (solid circles) along with the ac-susceptibility (lines). The inset shows the $T$ dependence of $^{71}$Ga-NQR frequency. }
\label{1/T1TvsT}
\end{figure}

Figure~2 shows 1/$T_1T$ as a function of $T$ in the temperature range of $1 \leq T \leq 280$\,K along with the ac-susceptibility measured using an in-situ NQR coil. 
In the normal state above $T_{\rm c}$, a $T_1T=$\,const. relation is valid up to $T \sim 50$\,K, which is usually seen in conventional metals. Above $T \sim 50$\,K, although 1/$T_1T$ gradually deviates from the constant, it is probably ascribed to the local structural distortion of the crystal because NQR frequency also shifts in the same $T$ range, as seen in the inset of Fig.~2. This structural change may bring about an increase in the density of states at the Fermi level, leading to an increase of the $1/T_1T$ constant value in the high $T$ region. It is important to investigate the precise crystal structure of Ir$_2$Ga$_9$ at a low $T$ where the superconductivity emerges.
  
In the superconducting state, NQR intensity is largely suppressed below $T_{\rm c}=2.2$\,K, as shown in the inset of Fig.~1. This is consistent with the previous reports that Ir$_2$Ga$_9$ is nearly a type-I superconductor \cite{Shibayama,Wakui}, which prevents the penetration of the rf field into the sample due to the Meissner effect.

Figure~3 shows the $T$ dependence of $1/T_1$, focusing on the superconducting state.
It should be noted that $1/T_1$ shows a distinct coherence peak just below $T_{\rm c}$, and decreases exponentially upon further cooling. This is typical behavior of a conventional $s$-wave superconductor. Generally, $1/T_1$ in the superconducting state is expressed as
\[\displaystyle \frac{1}{T_1}\propto \int^{\infty}_0 dE \, [N_{\rm s}^2(E)+ M_{\rm s}^2(E)] f(E)[1-f(E)],\]
where $N_{\rm s}(E)=N_0E/(E^2-\Delta^2)^{1/2}$ is the density of states (DOS) in the superconducting state, $M_{\rm s}(E)=N_0\Delta/(E^2-\Delta^2)^{1/2}$ is the anomalous DOS induced by the coherence effect, $N_0$ is the DOS at Fermi level in the normal state, and $f(E)$ is the Fermi distribution function \cite{Hebel1}.
When a superconducting energy gap $\Delta$ is isotropic, both $N_{\rm s}(E)$ and $M_{\rm s}(E)$ diverge at $E=\Delta$, which yields the coherence peak just below $T_{\rm c}$. Because of the inevitable damping effect of quasiparticles in real materials, here, $N_{\rm s}(E)$ and $M_{\rm s}(E)$ are averaged over an energy broadening function, assuming a rectangle shape with a width 2$\delta$ and a height 1/2$\delta$ \cite{Hebel2}. 
In Fig.~3, it is observed that a theoretical curve based on an isotropic $s$-wave model is actually in good agreement with the experiment using $2\Delta(0)/k_{\rm B}T_{\rm c}=4.4$ and $\delta/\Delta=0.25$, as shown by the solid curve. This indicates that Ir$_2$Ga$_9$ is a conventional $s$-wave superconductor. Although there is no spatial centrosymmetry in the crystal, no unconventional behavior due to ASOC  emerges in the superconducting state.
 
It is remarkable that the magnitude of the coherence peak is significantly large, increasing at 2.04\,K up to $\sim 1.5$ times larger than that at $T_{\rm c}=2.2$\,K. The anisotropy of the Fermi surface and the pairing interaction suppress the coherence peak in general. In contrast, the coherence peak is often clearly observed in noncentrosymmetric compounds with no strong electron correlation such as Li$_2$Pd$_3$B \cite{Nishiyama2}, LaIrSi$_3$ \cite{Mukuda}, and La$_2$C$_3$. It is suggested that the isotropy of the Fermi surface would be improved by the splitting of Fermi surface due to ASOC. 
Further experiments to address the noncentrosymmetric effect on the coherence of $\Delta$ are required.

In conclusion, the $^{71}$Ga-NQR measurements have revealed that $1/T_1$ has a clear coherence peak just below $T_{\rm c}$ and decreases exponentially upon further cooling in Ir$_2$Ga$_9$. From these results, Ir$_2$Ga$_9$ is concluded to be a conventional $s$-wave superconductor.  Despite the lack of spatial centrosymmetry, there is no evidence of the unconventional superconducting state ascribed to ASOC in Ir$_2$Ga$_9$. This may be because the energy scale of ASOC $E_{\rm so}\sim 200$\,meV is much smaller than the Fermi energy $E_{\rm F}\sim12$\,eV for Ir$_2$Ga$_9$ ($E_{\rm so}/E_{\rm F}\sim 0.017$) \cite{Bostrom,Shibayama}. However, $E_{\rm so}/E_{\rm F}\sim 0.03$ is also small for Li$_2$Pt$_3$B \cite{Fujimoto}. It remains an underlying issue why Li$_2$Pt$_3$B is unconventional only owing to ASOC in superconductors without strong electron correlation.  

This work was supported by a Grant-in-Aid for Creative Scientific Researchi15GS0213), MEXT and the 21st Century COE Program supported by the Japan Society for the Promotion of Science.

\begin{figure}[h]
\centering
\includegraphics[width=7.0cm]{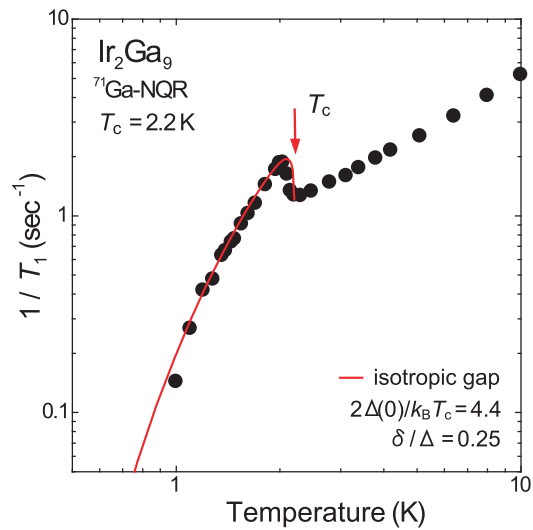}
\caption[]{(Color online) Temperature dependence of $1/T_1$. The solid line is the calculation based on an isotropic SC gap with $2\Delta(0)/k_{\rm B}T_{\rm c}=4.4$ and $\delta/\Delta=0.25$.   }
\label{1/T1vsT}
\end{figure}



\end{document}